\documentstyle[11pt,epsf]{article}
\def\be{\begin{equation}}
\def\ee{\end{equation}}
\begin{document}
\baselineskip=18pt
\vspace{1cm}

\begin{center}
Comments on: ``Weinberg's  Nonlinear  Quantum  Mechanics
and Einstein-Podolsky-Rosen paradox", by Joseph Polchinski. 
\vskip1cm

Bogdan Mielnik${}^{a, b}$ 
\vskip0.5cm
{\it
${}^a$Departamento de F\'{\i}sica, CINVESTAV, AP 14-740\\
07000 M\'exico DF, M\'exico; e-mail: bogdan@fis.cinvestav.mx\\[2ex]
${}^b$ Institute of Theoretical Physics, Warsaw University,\\  
Hoza 69, Warsaw, Poland
}
\end{center}
\vskip1cm

\begin{abstract}
\baselineskip=18pt
Contrary to the opinion of J.  Polchinski  [Phys.Rev.Lett.  {\bf 66},  
397-400 (1991)], the phenomenon of superluminal messages in nonlinear
versions of quantum mechanics is not a specific difficulty in a class of
theories formulated by S. Weinberg [Ann.Phys. (N.Y.), {\bf 194}, 336-386
(1989)]. It appears in all schemes which try to enlarge the orthodox class
of observables, while conserving the traditional structure of the pure
and mixed states. 
\end{abstract}

In the last decades some attention was dedicated to the  
cases of quantum mechanics (QM) based on nonlinear wave equations. One of 
most elegant attempts was presented by S. Weinberg \cite{Wei} by applying
the 
Hamiltonian formalism to the complex wave functions.  Soon  however,  it 
was shown, that the  scheme  when  applied  to  many  particle  systems, 
generates instantaneous  messages  between  distant  components  in  the 
measurements  of  Einstein-Podolsky-Rosen  (EPR)  type  (N.Gisin
\cite{Gis}, 
M.Czachor \cite{Cza}). It was henceforth concluded that the nonlinear
QM contradicts the causality. The conclusion has 
been amended by J.Polchinski \cite{Pol}, who has argued that  the
superluminal  effects are  just a special fault of Weinberg's
formalism but can be avoided  in  a wider class of nonlinear theories.
Since that time, the idea seems accepted (see, e.g. \cite{cza}) without 
any fundamental critiques.  Below, I show that the argument of 
Polchinski fails: neither the difficulty is specific to  the  
Weinberg's scheme, nor the recipe offered in \cite{Pol} permits 
to obtain new types of causal but nonlinear QM. 

As it seems, the atypical variants of QM  may break with the
orthodox scheme in several ways, e.g.: 
(I) They can  modify  the manifold of 
pure states; (II) they can adopt the  orthodox  (linear)  space of pure
states but assume the  existence  of  nonlinear  evolution  operations;
(III) they can adopt the orthodox manifold  of  pure  states  but modify
the class of the  {\it functional  observables}  (with  or without 
introducing the nonlinear evolution).   

Since the criterion of Polchinski refers  to  the  observables, we shall  
discuss (III). Consider a pair of hypothetical quantum systems A  and  B 
with the pure states described by the unit spheres $S_A$, $S_B$ in two 
Hilbert  spaces $H_A$ and $H_B$. Following \cite{Gis,Pol}, we neglect
the motion of  both objects;  so $H_A$ and $H_B$ represent the internal
degrees.  We then adopt the tensor product
space $H_A \otimes H_B$ to describe 
the entangled system (all vectors in $H_A \otimes H_B$ 
represent  the  admissible  states, the simple products  
$a \otimes b = |a> |b>$  mean  no  correlation).  
We also take for granted that all traditional 
measurements, represented by the orthogonal projectors in 
$H_A$, $H_B$ can be performed on the single components of 
the entangled pair. (We have adopted some essential 
elements of the orthodox structure, to share the partition point 
with \cite{Gis,Cza,Wei,Pol}). In addition, we assume that one 
of  the  systems,  e.g. B, is atypical in the sense (III), permitting to
measure  at  least one observable $f : S_B \longrightarrow \Re$ which 
might not be a quadratic form on $S_B$. 

Following EPR, let us now imagine a source which produces a 
sequence  of identical, entangled states: 
\be
\Psi = \alpha_1 |A_1> |b_1> +...+ \alpha_n |A_n> |b_n>   
\label{1}
\ee
bombarding (with a fixed frequency) two 
distant observers,  `Alice' and 
`Bob'. Alice obtains A-objects; she tries to affect the entangled system 
at her end by performing measurements on $|A>$ states; Bob will try to use 
$f$  to  read  the  Alice  doing.  Since  the  Alice  measurements  reduce 
$|A>$-states to orthogonal systems, we  loose little  by  assuming
that $|A_i>$ ($i=1,...n$) are orthonormal. We don't assume the same about
$|b_i>$'s, but  only that  $<b_i|b_i>=1$ ($i=1,...n$) and $|\alpha_1|^2
+...+|\alpha_n|^2 = 1$. Thus,  all simple products in (\ref{1}) are 
mutually orthogonal in $H_A \otimes H_B$ and $<\Psi|\Psi>=1$. Suppose
that Alice measures an observable A with (nondegenerate) eigenvalues
$\lambda_i$ on eigenstates $|A_i>$ ($i=1,...n$). If she obtains
$\lambda_i$, Bob `receives' the  pure state $|b_i>$ (note, that the
justification does not necessarily involves  the v.Neumann  projection
postulate  applied  in  $H_A \otimes H_B$;  as long as our theory 
includes the traditional measurements on each subsystem, the 
strict statistical correlation on  both  ends  gives  as  
credible  argument;  compare  the  `teleportation' 
\cite{Ben}).  Thus,  if  Alice  performs  a  sequence  of 
A measurements on her side, Bob will receive a random sequence $b =
{|b_1>, |b_2>, |b_3>,...}$, each $|b_i>$ repeating itself with  the
frequency $p_i = |\alpha_i|^2$ . Suppose now,  Alice switched 
to a new apparatus $A'$ with new 
(orthonormal) eigenstates $|A_1'>,...,|A_n'>$  ($|A_j'>$ and $|A_i>$  
spanning the same subspace of $H_A$). 
The  entangled state (\ref{1}) admits
an alternative expression: 
\be
\Psi = \alpha_1' |A_1'> |b_1'> +...+ \alpha_n' |A_n'> |b_n'>
\label{2} 
\ee
where the unit vectors $|b_i'>$ and the coefficients $\alpha_i'$ 
can be easily calculated. Now, if Alice measures A$'$,  
Bob receives a new  sequence of states  
$b' = {|b_1'>, |b_2'>,...}$ appearing with 
the new frequencies $p_i'=|\alpha_i'|^2$ . Since the single states 
are not recognizable,  the  entire sequencies $b$ and $b'$ must be the
`letters' of Alice alphabet. Can Bob read  them?  To  distinguish  $b$ 
and $b'$  he  has  the  conventional observables  (of no use!), but he can
apply also  the  observable  $f$.  By  measuring $f$ on $b$, he finds the
statistical average: 
\be
f[b] = p_1 f(|b_1>) +...+ p_n f(|b_n>),                
\label{3}
\ee
while for $b'$ he obtains: 
\be
f[b'] = p_1' f(|b_1'>) +...+ p_n' f(|b_n'>).
\label{4}
\ee

These averages are also considered by Weinberg,  though  questioned 
by Polchinski. Yet, there is some {\it quid pro quo} in \cite{Pol}, 
almost like in Esher's drawings \cite{Esh}.  Indeed, if one  
does not insist on the orthodox scheme of `operator observables' 
\cite{Dir}, then the observables are just c-number 
functions  on  states, representing the statistical averages
\cite{Mie, Haa, Kib, Wei}. It means that some universal facts 
concerning the statistical ensembles must be valid. 
If an ignorant observer measures $f$ for a sequence 
of randomly received states, without knowing 
{\em which is which}, he must 
unavoidably find the statistical averages (\ref{3}-\ref{4}). 
Thus, (\ref{3}-\ref{4}) have a universal validity.  
(At  least, nothing can stop 
Bob from making precisely this  statistics 
at his end!).  The  concept  of  a  "density  matrix", 
meanwhile, is particular; in fact, it turns insufficient to describe 
the mixed states in nonlinear theories  \cite{Mie, Haa, Kib}. 
What Polchinski assumes is that the density matrices of the 
$B$-subsystem, still contain enough 
information to determine the values (\ref{3}-\ref{4}) for  
the observable $f$ on the $b$-sequencies. If one 
adopts the idea,  the rest of the story develops in $H_B$.  
If $b$ and $b'$ are `generated' by Alice (by measuring 
A and $A'$), then the simple calculation shows that 
their `density matrices' coincide: 
\be
\rho = \sum |\alpha_i|^2 |b_i><b_i| = 
\sum |\alpha_i'|^2 |b_j'><b_k'| = \rho ' 
\label{5}
\ee
The criterion \cite{Pol} then says $f[b] =f[b']$; 
so $f$ does not distinguish Alice letters.
One might hope that by choosing arbitrary $f(\rho)$  
one can arrive at distinct no-signal theories, but this turns 
out an illusion. The point is  that an observable (statistical 
average) cannot be postulated without caring for the 
consistency conditions, which interrelate its values on the mixture 
with the values on the mixture components. As a consequence, if 
(\ref{3}) and (\ref{4}) coincide for any two sequencies  
generated by Alice (yielding a well defined function of $\rho$),   
then $f$ can be only a quadratic form on $S_B$. 
To illustrate this, take dim $H_B = 2$. The convex 
set of all density matrices in $H_B$ can be represented as the 
unit ball $R_1$ in $\Re^3$. The ball surface $S^2$, (i.e., the 
projective unit sphere in $H_B$) collects the simple density 
matrices of the form $|b><b|$ (rays in $H_B$). We stick to the assumption
that they  represent the pure states of the $B$-subsystem. The antipodal 
points of $S^2$ stand for orthogonal rays. The "density matrices" in 
$H_B$ are arbitrary points $x \in R_1$  (Fig.1);  the convex linear 
combinations $p_1 x_1 + p_2 x_2$ ($p_1, p_2 \geq 0$, $p_1 + p_2 = 1$) 
for $x_1, x_2 \in R_1$ define the natural geometry of 
$R_1$ \cite{Mie}. We adopt the idea \cite{Pol} that they contain 
some (at least partial) information 
about the physical mixtures and that $p_1,p_2$ are 
the mixing probabilities. Consider now two pairs of 
points (pure states) $x_1$, $x_2$ and $x_1'$, $x_2'$ on $S^2$. Following
\cite{Pol}, we assume that if the straight line intervals $x_1 x_2$  
and $x_1' x_2'$ intersect at a point  
$x=p_1 x_1 + p_2 x_2 = p_1' x_1' + p_2' x_2' \in R_1$, the 
values of $f$ on both mixtures must coincide: 
\be
p_1 f(x_1) + p_2 f(x_2) = p_1' f(x_1') + p_2' f(x_2').           
\label{6}
\ee

so that (\ref{6}) becomes a well defined function $\Phi (x)$ of 
$x = p_1 x_1 + p_2 x_2 \in R_1$ : 
\be
p_1 f(x_1) + p_2 f(x_2) = \Phi (p_1 x_1 + p_2 x_2).
\label{7}
\ee

\begin{figure}[htbp]
\begin{minipage}{6truecm}
\hspace*{3.7truecm}
\epsfxsize=7truecm
\epsfbox{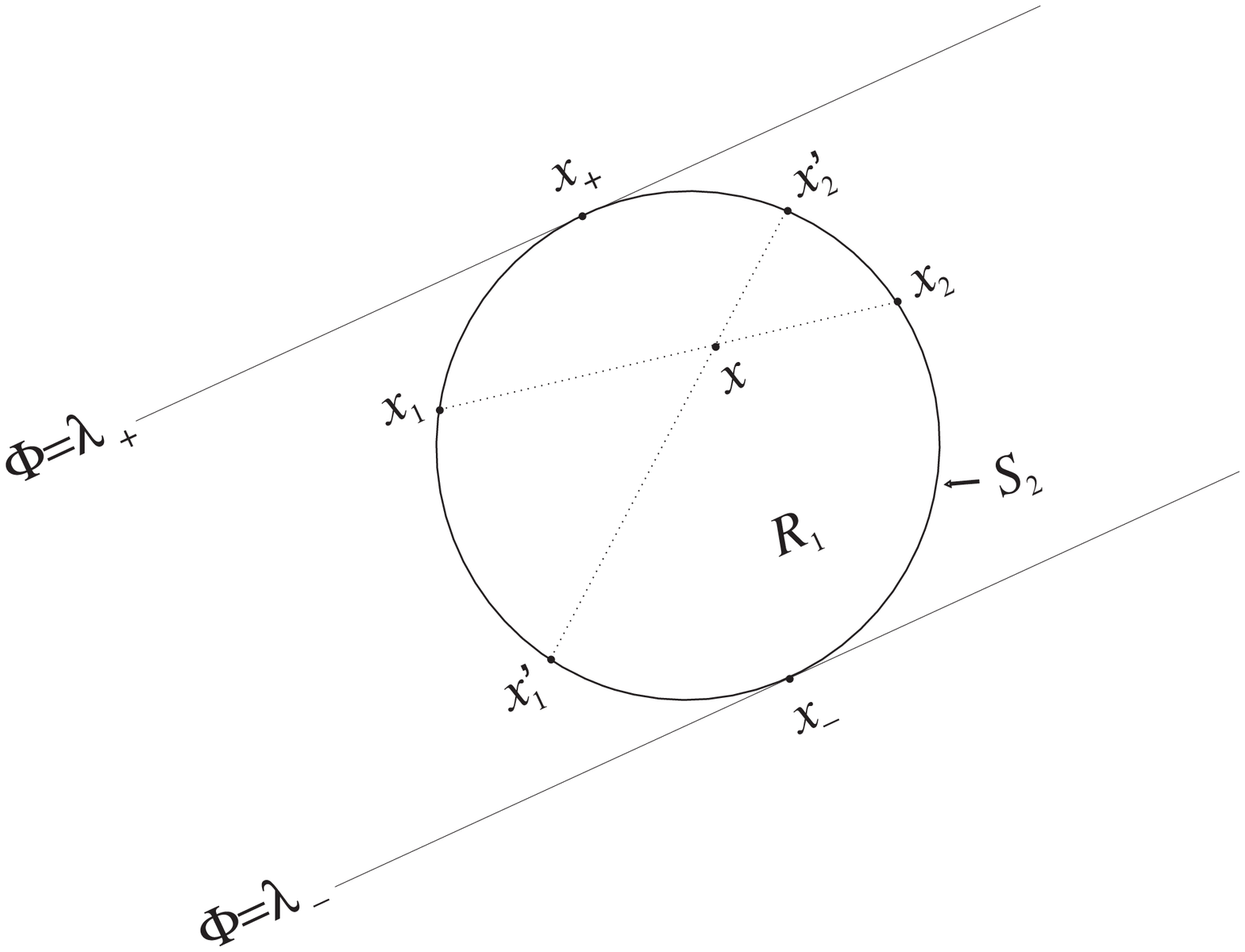}
\end{minipage}
\bigskip
{\caption{\small 
Due to the natural geometry of the density
matrices in ${\rm dim} \ H_B=2$, the 'no-signal condition'
of Polchinski can be satisfied only by the affine
functionals in $\Re^3$, corresponding to the quadratic
observables $f:S_B \longrightarrow \Re$.}}
\end{figure}

By physical arguments, $\Phi$ should be continuous. Putting $p_1 = 1$,
$p_2 = 0$ or $p_1 = 0$, $p_2 = 1$, one gets $f(x_1) = \Phi (x_1)$ and
$f(x_2) = \Phi (x_2)$, which converts (\ref{7}) into: 
\be
\Phi (p_1 x_1 + p_2 x_2) = p_1 \Phi (x_1) + p_2 \Phi (x_2),
\label{8}
\ee
i.e., $\Phi$ is linear with respect to the convex combination in $R_1$.
Since $R_1$ spans $\Re^3$ , it is the matter of simple extension to
consider $\Phi$ linear on $\Re^3$ with respect to the {\it affine linear
combination} $p_1 x_1 + p_2 x_2$,  ($p_1, p_2 \in \Re$, $p_1 + p_2 = 1$).
If $\Phi \neq$ const in $\Re^3$,  then  the  equations $\Phi =$ const
determine a congruence of closed, parallel planes in $\Re^3$. Two of them
are tangent  to $S^2 = \partial R_1$ in two antipodal points
$x_{\pm} = |b_{\pm}><b_{\pm}|$ where $\Phi$ accepts its  maximal 
and minimal values $\lambda_{\pm}$ (on $R_1$). Exactly  the  same
properties has the functional $\lambda_+ <b_+| x |b_+> + 
\lambda_-<b_-| x |b_->$. Thus: $\Phi (x) = \lambda_+ 
<b_+| x |b_+> +  \lambda_- <b_-| x |b_->$. In particular, for $x
=|\psi><\psi| \in S^2$ , $\Phi (x) = \lambda_+ |<b_+|\psi>|^2 +
\lambda_- |<b_-|\psi>|^2$, i.e., $\Phi$ is just a quadratic form of the
pure states $\psi$. If $\Phi$ is constant in $\Re^3$, the same holds with
$\lambda_+ = \lambda_-$. 

Paradoxically, the proof is even simpler if dim $H_B \geq 3$. 
Let us recall that all quantum measurements can be reduced 
to elementary `counting experiments' (carried out  by unsophisticated 
counters which can either detect or overlook the particle). If $f$ is
a `counting observable', then  $0 \leq f[b] \leq 1$. Suppose, $f$
satisfies the condition  of Polchinski \cite{Pol}. Let $X \subset H_B$ be
a subspace (dim $X = n$), $P_X$ the  corresponding projector and
$|b_1>, ..., |b_n>$ any orthonormal basis in $X$;  then  the  sum 
$f[b] = \left(\frac{1}{n}\right)\left[f(|b_1>) +... + f(|b_n>)\right]$ 
does not depend on the basis, but only on the entire subspace $X$. The
same concerns the `renormalized' sum: 
\be
f(|b_1>) +... + f(|b_n>) = n f[b] = \mu (X) 
\label{9}
\ee
which therefore defines a non-negative measure $\mu$ on the subspaces
$X \subset H_B$. By taking two subsequencies $|b_1>, ..., |b_r>$ and
$|b_{r+1}>, ..., |b_n>$, and the corresponding two orthogonal subspaces
$Y$, $Z \subset H_B$, $Y + Z = X$, $X \bot Y$, we  see  from 
(\ref{9}) that $\mu (Y) + \mu (Z) = \mu (X)$, i.e., $\mu$ is a positively
defined, orthoadditive measure on the subspaces $X \subset H_B$.  Since
dim $H_B \geq 3$, Gleason  theorem \cite{Gle} implies the existence of a  
non-negative operator $F : H_b \rightarrow H_b$,  such  that 
for any $X \subset H_b : \mu (X) = Tr(FP_X)$,  where  $P_X$  are  the
orthogonal projectors associated with the subspaces $X \subset H_B$. In
particular, if $X$ is a 1-dim subspace spanned by the unit vector
$|\psi>$, and  $P_X =|\psi><\psi|$,  then  $f(|\psi>) =
\mu (X) = Tr(F|\psi><\psi|) =\\ <\psi|F|\psi>$; i.e., $f$ is a quadratic
form on $S_B$. (Notice, that we have slightly strengthened the original 
Gisin  argument \cite{Gis}, by limiting the Polchinski condition for 
$f$ to the orthogonal $b$-sequencies). Since any observable is a linear
combination of `counting observables', we have shown that any
observable satisfying the Polchinski criterion must be quadratic on 
$S_B$. The non-quadratic  observables  protected 
against the superluminal effects are an  illusion (just give me 
one non-quadratic form on $S_B$, representing a statistical
average, and nothing can stop me from using (\ref{3}-\ref{4}) 
to read Alice messages!). 

We conclude  that  the  superluminal  effects  are  not  a specific 
difficulty of Weinberg's approach, but a generic phenomenon in nonlinear 
theories which have absorbed too ample fragments of the orthodox scheme. 
A way out, perhaps, could be a consistent deformation of the pure and 
mixed states, as well as the functional observables. 
This is, however, a different story which still waits to be written.

\end{document}